# Two-dimensional metalorganic ferromagnets

*Egzona Isufi Neziri,[a] Céline Hensky,[b,c] Hien Quy Le,[b,c] Diego Radillo Ochoa,[b,c] Aleksandra Cebrat,[a] Manfred Parschau,[a] Karl-Heinz Ernst[\*,a,d,e] Christian Wäckerlin[\*,b,c]*

[a] Molecular Surface Science Group, Empa, Swiss Federal Laboratories for Materials Science and Technology 8600 Dübendorf (Switzerland)

[b] Institute of Physics, Swiss Federal Institute of Technology Lausanne (EPFL) Station 3 CH-1015 Lausanne (Switzerland)

[c] Laboratory for X-ray Nanoscience and Technologies, Paul-Scherrer-Institut (PSI) CH-5232 Villigen PSI (Switzerland)

[d] Nanosurf Laboratory, Institute of Physics, The Czech Academy of Sciences 16200 Prague (Czech Republic)

[e] Department of Chemistry, University of Zurich, CH-8057 Zürich (Switzerland)

KEYWORDS: 2D metalorganic magnet, Scanning Probe Microscopy, X-ray Magnetic Circular Dichroism

ABSTRACT: Driven by applications in information technology, the search for new materials with stable, long-range magnetic ordering continues. Metalorganic magnets, involving the coordination of metal atoms with specific organic ligands, are a focus of intense research. These magnets offer customizable properties through synthetic adjustments to ligands or coordination chemistry. Here the synthesis, structural characterization, and magnetic properties of the 2D cyanocarbon magnet NiTCNE is reported. 2D-crystalline domains of this single-layered metalorganic network reach sizes exceeding 30 nanometers through co-deposition of the ligand TCNE (tetracyanoethylene) and Ni atoms on an Au(111) surface under ultrahigh vacuum conditions. Non-contact atomic force microscopy visualizes the structure with atomic resolution. X-ray magnetic circular dichroism establishes the 2D NiTCNE as a ferromagnet, with a very high magnetic remanence, a coercive field of around 1 tesla and a Curie temperature between 10 and 20 kelvin. As metalorganic chemistry opens a large variety of routes of synthesis, we anticipate that this materials research paves the way to new magnetic nanomaterials for spintronics applications.

Single-layer metalorganic magnets[1,2] offer promising applications as magnetic semiconductors[3] and as model-systems to rationalize and develop future quantum materials. As single layers require only tiny amounts of substance, they render these materials extremely resource efficient. Reports on ferromagnetism in low-dimensional materials, such as organometallic lanthanide compounds,[4] van-der-Waals materials such as $CrI_3$[5] as well as stacked and single-layer metalorganics,[6–8] and organometallics showcase the possibility to establish long range magnetic ordering in low-dimensional hybrid materials.

In bulk form, the complexes of the strong electron acceptors tetracyanoquinodimethane (TCNQ) and the structurally similar but smaller tetracyanoethylene, TCNE (Figure 1a) were the foundation for collective magnetism in organic and metal-organic materials.[9,10] The ferromagnetic cyanocarbon $V(TCNE)_2$, discovered more than 30 years ago, features the highest estimated characteristic temperature for any coordination compound in the solid state to date.[11] Based on this observation, and inspired by the fact that 2D NiTCNQ is superparamagnetic,[12–14] the present report establishes single-layer cyanocarbon material NiTCNE as a true ferromagnet. Its atomic-level structure is resolved with scanning probe microscopy (SPM) and the chemical states of NiTCNE are characterized by in-situ X-ray photoelectron spectroscopy (XPS). The magnetic properties are determined by X-ray absorption spectroscopy (XAS) and X-ray magnetic circular dichroism (XMCD) measurements.

**Results and Discussion**

2D NiTCNE is fabricated by on-surface synthesis in ultrahigh vacuum (UHV) on an inert Au(111) substrate. Specifically, well-ordered 2D crystals of NiTCNE are produced by codeposition of Ni atoms and TCNE molecules on the substrate kept at slightly elevated temperatures (see methods section for details).

*Chemical Characterization by X-ray Photoelectron Spectroscopy*

2D NiTCNE has C 1s and N 1s signals with binding energies of 285.6 eV and 398.7 eV, respectively, and has its Ni $2p_{3/2}$ signal at ~ 853.2 eV (Figure 1b-d). Compared to pure TCNE, the N 1s and C 1s peaks are shifted towards lower binding energies. Conversely, the Ni $2p_{3/2}$ signal is shifted to higher binding energies compared to metallic Ni. The shift to higher binding energy for nickel and to lower binding energies for carbon and nitrogen implies charge transfer from the metal atoms to TCNE, consistent with data obtained for its bulk analogue.[15,16]

*Molecular Structure Revealed by Scanning Probe Microscopy*

In overview STM scans (Figure 1e,f), well-ordered domains exhibiting a square lattice extending tens of nanometers are observed. The good 2D crystalline order is confirmed by sharp peaks in the Fourier-transform of the STM data (Figure S1). In constant height STM mode with a carbon monoxide modified tip, the TCNE molecules are imaged as dumbbell-shaped entities (Figure 1g). The Ni atoms are imaged as round protrusions (Figure 1g). The concurrently performed atomic force microscopy (AFM) is highly sensitive to Pauli repulsion.[17,18] Thus, the frequency shift image (Figure 1h) reveals well the molecular structure of TCNE. The visibility of the molecular backbone is further enhanced in the Laplace filtered version of the AFM data (Figure 1j). The nickel atoms are not observed in the AFM contrast (Figure 1h,i) as bright protrusions because they are located a fraction of an angstrom below the plane defined by the TCNE molecules or because they are slightly positively charged, thereby interacting less repulsive with the CO probe particle.

The analysis of the SPM data reveals that the nickel atoms are arranged in an essentially square lattice (Ni − Ni distance: 0.67 nm) by coordination with four cyano-groups from four TCNE molecules each. However, although TCNE has a two-fold $C_2$ axis, it adopts two 90°-rotated orientations at seemingly random positions without apparent distortion of the nickel lattice (Figure 1g).

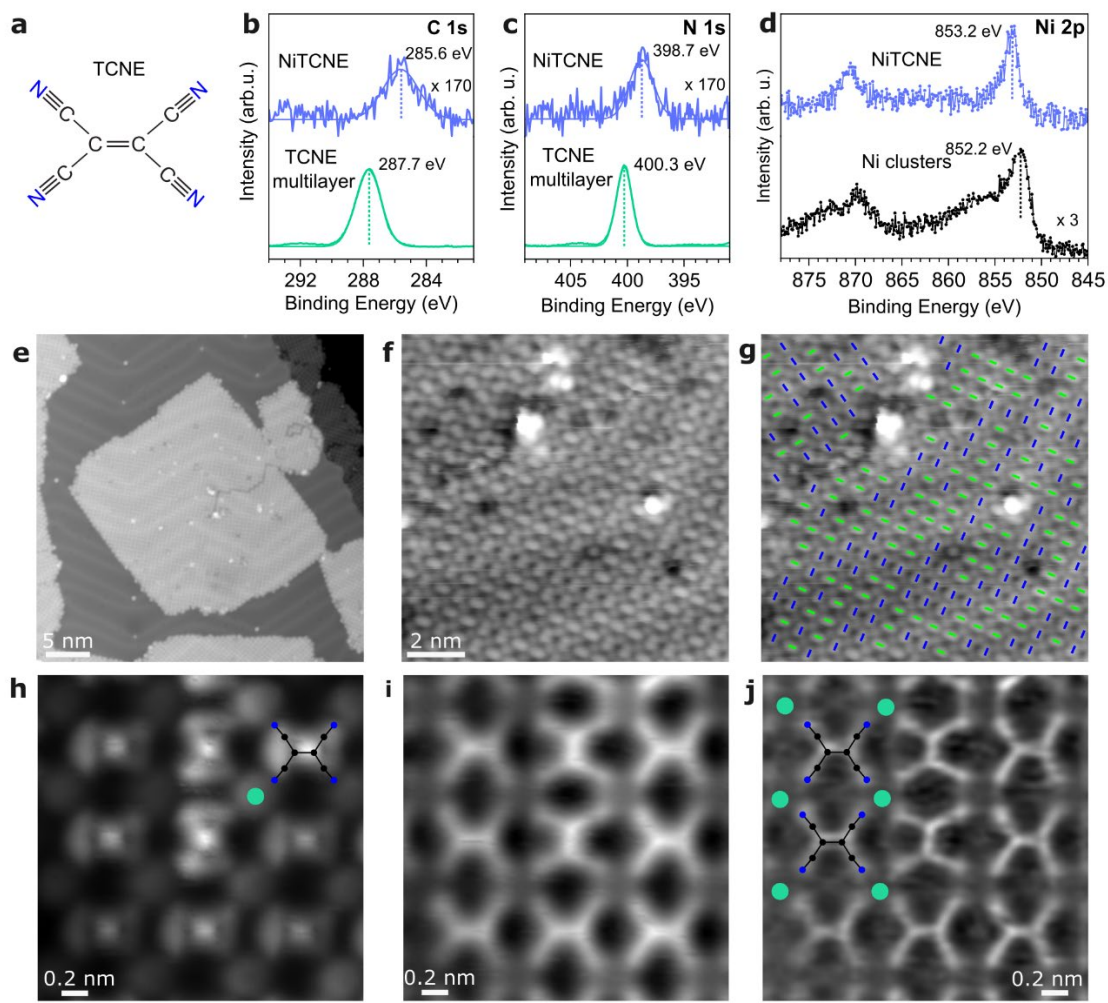

**Figure 1.** XPS and SPM data of the 2D NiTCNE on Au(111). (a) Structure of TCNE. (b-d) C 1s, N 1s and Ni 2p XPS of NiTCNE compared with (b, c) a multilayer of TNCE molecules and (d) metallic Ni clusters on Au(111). The C 1s and N 1s binding energies of NiTCNE are lowered compared to the one of the TCNE reference while the Ni $2p_{3/2}$ binding energy is increased with respect to metallic Ni. The chemical shifts univocally demonstrate the charge transfer from Ni to TCNE. (e-j) Overview STM images (e-g) and constant height STM (h) / AFM (i) scans recorded with a carbon monoxide functionalized probe. Laplace filtered version of the AFM image (j) to aid with the identification of the molecular structure. Imaging parameters reported in Table S1. Well-ordered layers of NiTCNE are observed. Within the essentially square lattice defined by the Ni atoms ($d_{Ni-Ni}$ = 0.67 nm), TCNE adopts two orientations (annotated in g with blue and green lines). The Ni atoms are best seen in the STM contrast (h) as round protrusions (annotated as green disks). Four TCNE molecules coordinate one Ni atom with their cyano-groups, best seen in the AFM data (i, j).

*Magnetic Properties Determined by X-ray Magnetic Circular Dichroism*

X-ray magnetic circular dichroism (XMCD), which is the difference in X-ray absorption spectroscopy (XAS) for right and left circularly polarized light ($\sigma_+$ and $\sigma_-$), is a direct measure of the magnetization projected on the beam direction. As XAS/XMCD is highly surface sensitive in the chosen set-up, it is perfectly suited to unravel the magnetic properties of magnetic systems. Figure 2 presents XAS/XMCD data of NiTCNE on Au(111), obtained at the Ni $L_{3,2}$ edges at 3 K. The experimental geometry and X-ray polarization modes are sketched in Figure 2a. In these experiments, the external magnetic field is co-linear with the direction of X-ray incidence. Hence, the XMCD data recorded in normal ($\theta = 0°$, out-of-plane) and grazing ($\theta = 60°$, approximately in-plane) incidence allows determination of the magnetization of the layer in these two directions.

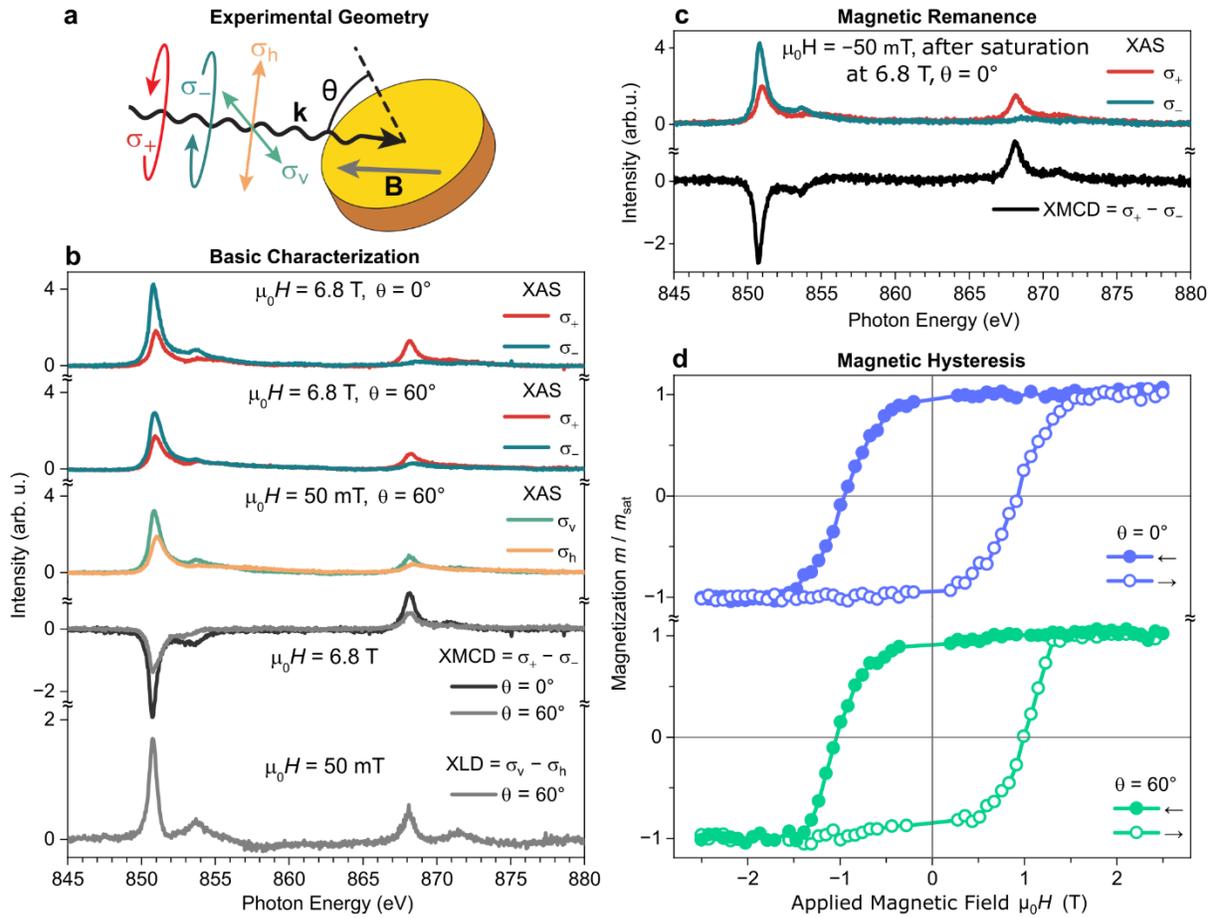

**Figure 2.** XAS/XMCD/XLD characterization of NiTCNE on Au(111), magnetic remanence and open magnetic hysteresis. Data recorded at a sample temperature of 3 K on the Ni $L_{3,2}$ edges. (a) Sketch of the experimental geometry and definition of the X-ray polarization: $\sigma_+$ and $\sigma_-$ denote spectra recorded with circularly polarized light; $\sigma_v$ and $\sigma_h$ correspond to linearly polarized light. $\theta$ denotes the x-ray incidence angle with respect to the surface normal. The applied magnetic field is co-linear with the direction of X-ray incidence. (b) XAS/XMCD at high magnetic field (magnetic saturation) and XAS/XLD that identifies structural ordering and is the fingerprint of the crystal field acting on the Ni atoms. (c) XMCD recorded at –50 mT after magnetic saturation at 6.8 T demonstrates remanent magnetization with unchanged sign. (d) XMCD magnetization

curves (sweep rate: 0.75 T/min; "←": sweep from 2.5 T to –2.5 T; "→": sweep from –2.5 T to 2.5 T) revealing magnetic hysteresis.

XAS/XMCD data recorded in normal and grazing incidence at high magnetic field (6.8 T) establishes the saturation magnetization (Figure 2b). The XAS/X-ray linear dichroism (XLD) data also shown in Figure 2b is a fingerprint of the charge distribution around the Ni atom. Based on these three sets of data, multiplet calculations using the code multiX[19] (Figure S2) univocally demonstrate that nickel is in the +1 oxidation state, *i.e.* it has a $3d^9$ electronic configuration. As described in the methods section, the multiplet calculations also permit to determine the magnetic spin moment $<S_z>$ from the experimentally measurable value $<S_{z,eff}>$ that is obtained by the sum-rule analysis. The total magnetic moments $m_{tot}$ of NiTCNE at saturation (6.8 T and 3 K) in normal and grazing incidence are identical within the limits of uncertainty (Table 1). This confirms that NiTCNE is magnetically fully saturated in both the out-of-plane and the approximately in-plane direction. In other words, the magnetization has achieved full alignment in both directions. Hence, the magneto-crystalline anisotropy is small compared to the magnetostatic energy ($\sim 2 \times m_{tot} \times 6.8$ T $\approx 1.2$ meV).

*Magnetic Remanence*

Following the above-described basic characterization of the magnetic properties of NiTCNE at saturation, the magnetization out of equilibrium is studied. The foremost property of a magnet is its magnetic remanence, *i.e.,* the remaining magnetization in absence of an applied field. Figure 2c shows XAS/XMCD spectra recorded in normal incidence at –50 mT after saturating the magnetization at +6.8 T. A strong XMCD signal with an unchanged sign compared to saturation is observed. Comparison of the XMCD intensities shows that the remanent magnetization is 93% of the saturation magnetization. The slightly negative applied field excludes any possibility of a spurious remanence due to small residual positive field in the instrument that cannot completely ruled out by ramping the field back to 0 T.

*Magnetic Hysteresis*

The second important feature of a magnetic material is the presence of a hysteresis in the field dependent magnetization. Such magnetization curves are obtained by recording the field dependent XMCD signal in normal and grazing X-ray incidence (Figure 2d). Clearly, NiTCNE presents open hysteresis in both measurement geometries. The coercive fields, *i.e.*, external magnetic field that finally coerces the magnetization to change sign, are $H_c$ (0°) = 0.95 ± 0.02 T and $H_c$ (60°) = 1.01 ± 0.02 T for normal and grazing incidence, respectively (Table 1). The fact that $H_c$ (60°) is only slightly larger than $H_c$ (0°), implies a small out-of-plane magnetic anisotropy energy.[20] Conversely, if the magnetic anisotropy were large compared to the magnetostatic energy, e.g. in 3.5 ML Co/Au(332)[21] or Iron-9,10-dicyanoanthracene (DCA)/Au(111),[7] only the projection of the field on the magnetization direction is effective. In that case, $H_c$ (60°) can be up 2 $H_c$ (0°).

**Table 1.** Spin ($<S_z>$), orbital $<L_z>$ moments and total magnetic moments ($m_{tot}$) at saturation (6.8 T, 3 K) as well as coercive fields ($H_c$) of NiTCNE. $<S_{z,eff}>$ is the effective, uncorrected spin moment obtained by the sum rule analysis.

| $\mu_0 H$ \| T \| θ | $2<S_{z,eff}>$ [$\mu_B$] | $2<S_z>$ [$\mu_B$] | $<L_{z,eff}>$ [$\mu_B$] | $m_{tot} = (2<S_z> + <L_{z,eff}>)$ [$\mu_B$] | $\mu_0 H_c$ [T] |
|---|---|---|---|---|---|
| 6.8 T \| 3 K \| 0° | 1.19 ± 0.05 | 1.28 ± 0.06 | 0.14 ± 0.03 | 1.42 ± 0.07 | 0.95 ± 0.02 |

| | | | | | | |
|---|---|---|---|---|---|---|
| **6.8 T \| 3 K \| 60°** | 0.80 ± 0.08 | 1.46 ± 0.15 | 0.09 ± 0.05 | 1.60 ± 0.2 | | 1.01 ± 0.02 |

*Magnetic Ordering Temperature*

The magnetic ordering temperature is determined from temperature dependent magnetization curves (Figure 3). At 10 K, the hysteresis curve is still slightly open ($\mu_0 H_c = 55 \pm 30$ mT) but at 20 K no opening can be detected. Hence the critical temperature is in the interval 10 K < $T_c$ < 20 K. At 20 K and above, NiTCNE is superparamagnetic, *i.e.* the ferromagnetic interactions still favor parallel spin alignment, but they cannot stabilize magnetic ordering.

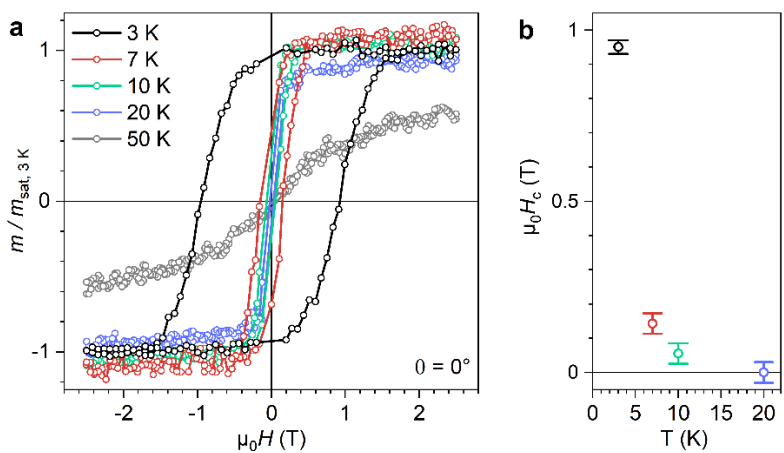

**Figure 3.** Temperature dependent magnetic hysteresis curves of NiTCNE/Au(111). (a) XMCD magnetization curves recorded in normal incidence ($\theta = 0°$) at a field sweep rate of 0.75 T/min. (b) Coercive fields $H_c$ as a function of temperature T. At 10 K, the hysteresis loop is still open (non-zero coercive field $H_c$). At 20 K there is no discernable hysteresis.

*Brief Summary of Supplementary Characterization*

N K edge XAS that probes the unoccupied density of states of TCNE (Figure S4) serves as supplementary characterization. The data shows a very strong polarization that implies the complete orientation of the TCNE molecules and is consistent with previous reports on NiTCNQ.[12] A further important result is that Ni impurities on Au(111) are completely non-magnetic (Figure S5 and S6). Therefore, although there is absolutely no indication for the presence of unreacted Ni impurities in the first place, the observed magnetic properties must originate exclusively from the NiTCNE layer. Finally, Figure S7 shows magnetization curves of NiTCNE/Au(111) recorded at 3 K over the complete field range (±6.8 T), confirming that 2D NiTCNE is completely magnetically saturated once the hysteresis opening closes at ~ ±1 T.

*Bipolar Magnetic Material*

Spin polarized density functional theory (DFT) calculations with Hubbard U term have been previously performed for NiTCNE in two polymorphs: i) a-NiTCNE where all TCNE molecules are aligned alike and ii) p-NiTCNE where every second TNCE molecule is rotated by 90°.[22] In addition to correctly predicting a $3d^9$ electronic configuration of Ni and a ferromagnetic spin alignment, it is found that NiTCNE is either a bipolar magnetic semiconductor (a-NiTCNE) or a bipolar magnetic metal (p-NiTCNE). In both cases, *bipolar* means that the density of states (DOS) above and below the Fermi level is fully polarized in the opposite spin-direction. This means that the application of a gate voltage or doping affects the magnetic properties and/or spin-transport. Intriguingly, the here experimentally realized 2D material contains both a- and p-polymorph arrangements in a non-periodic fashion (Figure 1g). Hence, its electronic properties can be naively expected to lie in between those predicted by theory. However, non-periodic arrangements are also expected to show novel, exotic properties in the (spin-resolved) density of states or (spin)-transport.[23] While an experimental validation of such predicted special electronic structure and (spin)-transport properties has to be yet realized, the fact that STM at a low bias voltage of 1.5 mV resolves the original lowest unoccupied molecular orbital (LUMO) of TCNE (Figure 1h) is fully in-line with the theoretical prediction that said orbital is essentially singly occupied and hence found at or very close to the Fermi level.[22,24]

*Exchange mechanism and comparison with existing single-layer low-dimensional metalorganics*

DFT theory predicts significant ferromagnetic coupling via orbitals of the TCNE ligand.[22] For the here experimentally realized NiTCNE on Au(111), a contribution from Ruderman–Kittel–Kasuya–Yosida (RKKY) interaction through electrons in the Au(111) substrate[25,26] cannot be firmly ruled out from an experimental point-of-view.

Compared to its direct relative NiTCNQ that is merely superparamagnetic,[12–14] NiTCNE is a ferromagnet with a high remanent magnetization and coercive field. Its Curie temperature and coercive field are lower than the ones reported for FeDCA[7] but higher than in case of EuCOT.[4]

**Conclusion**

It is actually surprising that single transition-metal TNCE layers have not been synthesized and tested for magnetism to date, despite the prototypical character of bulk TCNE organic magnets and the superparamagnetism in 2D NiTCNQ layers strongly suggested such investigations.[12,14] By establishing single-layer NiTCNE as a 2D cyanocarbon ferromagnet, our report rectifies now such previous shortcomings in terms of ferromagnetism. Specifically, 2D NiTCNE exhibits at 3 K both a remarkably high magnetic remanence of 93% of the saturation magnetization and a high coercive field of ~1 tesla. Open magnetic hysteresis is observed at least up to 10 K. XAS/XMCD/XLD, combined with multiplet calculations, reveal a $3d^9$ electronic configuration. Surprisingly, the impressive magnetic properties emerge *despite* the material having only a weak out-of-plane easy axis of magnetization. Scanning probe characterization reveals that NiTCNE forms a square 2D lattice (a = 0.67 nm) embedding two random orientations of the TCNE molecule in the lattice unit, rendering the 2D layer essentially non-periodic.

Based on DFT, 2D NiTCNE is expected to be a bipolar magnetic metal or semiconductor.[22] This implies that it should be possible to electrically control the magnetic properties as well as the spin-transport. Although the here presented characterization on a conductive Au(111) surface cannot unambiguously prove these predicted electronic properties and spin transport effects, the observation of a TCNE orbital essentially at the Fermi level is a strong indication that 2D NiTCNE is indeed bipolar. Furthermore, in view of the Mermin-Wagner theorem,[27] that formally forbids stable magnetic order in low dimensional systems without anisotropy, the remarkable magnetic properties observed despite low magnetic anisotropy offer new insights into low-dimensional magnetism. Finally, the not completely understood fact that the 2D ferromagnet can be magnetically saturated and switched at around 1 tesla regardless of the direction of the applied field is very advantageous for applications in spintronic devices. Low dimensional metal-organic magnetic have the advantages of requiring only small amounts of materials and that their properties can be tuned by choice and modification of the ligand molecule. We envision that they can be applied as active elements in information processing and storage devices, both classical and quantum.

**Methods**

*Sample Preparation and On-Surface Synthesis*

All experiments were conducted under ultra-high vacuum conditions (pressure p < $10^{-9}$ mbar). The Au(111) single crystal was cleaned by repeated cycles of Ar+ sputtering at 1.5 kV, followed by annealing to 500 °C. TCNE is a powder with sufficient vapor pressure to be sublimed by mild heating. Caution: TCNE forms toxic HCN with moisture from air. TCNE was dosed the vapor through a leak valve from a sealed and evacuated glass container kept at 60 °C after cleaning by pumping cycles with a turbomolecular pump.

The multilayer of TCNE on Au(111) was obtained by condensing TCNE (partial pressure 1 × $10^{-7}$ mbar) on the substrate kept at 110 K. 2D NiTCNE was obtained by depositing Ni atoms from a high temperature cell (Createc, $Al_2O_3$ crucible) at an operating temperature of 1150 °C in TCNE vapor on the sample kept at 70 °C. Subsequent annealing up to 100 °C was performed to enhance the long-range order of the MOF. Since TCNE desorbs from Au(111) at room temperature, the coverage of NiTCNE is solely determined by the amount of Ni sublimed. Unreacted TCNE desorbs. The coverages were calibrated by STM measurements.

*X-ray photoelectron spectroscopy*

XPS was measured using Al $K_\alpha$ X-rays (1486.7 eV), in normal emission with the sample kept at room temperature. The binding energy scale was calibrated using the Au $4f_{7/2}$ peak (84.0 eV) and the Fermi energy (0.0 eV).

*Scanning probe microscopy*

STM was measured using electrochemically etched and in-situ sputtered tungsten tips. Low temperature STM/AFM (Createc) was performed using a QPlus sensor with a PtIr tip. Imaging parameters are reported in Table S1. The room temperature STM images were obtained using the Aarhus 150 from Specs. At the synchrotron the samples were checked using the Omicron VTSTM.

*X-ray absorption spectroscopy*

XAS/XMCD/XLD experiments were conducted at the EPFL/PSI X-Treme beamline at the Swiss Light Source[28] in total electron yield mode. Circularly polarized ($\sigma_+$, $\sigma_-$) and linearly polarized ($\sigma_v$, $\sigma_h$) X-rays were used, with the magnetic field applied co-linear to the X-ray beam.

Background spectra recorded on the clean substrate were subtracted. The spectra were normalized to unity at the pre-edge. The XMCD and XLD spectra represent the differences ($\sigma_+ - \sigma_-$) and ($\sigma_v - \sigma_h$), respectively.

*Multiplet calculations*

The multiplet simulations were performed using multiX[19] for $3d^9$ and $3d^8$ electronic configurations using identical crystal-field parameters (see Table S2). The simulations based on the $3d^9$ electronic configuration reproduce the experimental data well (XAS, XMCD and XLD line shapes at $L_3$ and $L_2$ edges, presence of strong XMCD in both normal and grazing incidence). The simulations performed with a $3d^8$ electronic configuration fail to reproduce the experimental data.

*XMCD sum rule analysis*

The XMCD sum rule analysis[29–31] was performed with a number of holes $n_h = 1$, *i.e.* for the $3d^9$ electronic configuration yielding the effective spin and orbital moment projections $<S_{z,eff}>$ and $<L_z>$ (Figure S3). It is well known that the effective values obtained from the sum rules require correction.[29] Because at the Ni $L_{3,2}$ edges there is only very low mixing of $2p_{3/2}$ and $2p_{1/2}$ components, effectively only effect of the magnetic dipole operator $<T_z>$ needs to be corrected.[29,32] Since the used multiplet code multiX[19] considers both effects, the comparison of the sum-rule results applied to the simulated spectra ($<S_{z,eff}>$) with the calculated true values ($<S_z>$) directly yields the angle dependent correction factors $c_S = <S_{z,eff}>/<S_z>$ (Table S3) to determine the true spin moments from the experimental data.[33]

ASSOCIATED CONTENT

**Supporting Information**.

The following files are available free of charge.

SPM imaging parameters, 2D crystallinity of NiTCNE, Multiplet calculations and sum-rule analysis, Details on the sum-rule analysis and on the strongly anisotropic spin dipole moment, supplementary XAS/XMCD data (PDF)

**Data Availability Statement**

The data underlying this study are openly available in Zenodo.org at https://dx.doi.org/10.5281/zenodo.10977168.

AUTHOR INFORMATION

**Corresponding Authors**

* karl-heinz.ernst@empa.ch; christian.waeckerlin@epfl.ch

**Author Contributions**

The manuscript was written through contributions of all authors. All authors have given approval to the final version of the manuscript.

**Funding Sources**

Swiss National Science Foundation (SNSF) (202775)

Research Priority Program LightChEC of the University of Zürich

ACKNOWLEDGMENT

This work was supported by the Swiss National Science Foundation (SNSF) (202775) and by the Research Priority Program LightChEC of the University of Zürich. We thank J. Dreiser and V. Romankov for support during the synchrotron experiments.

Supporting Information

# Two-dimensional metalorganic ferromagnets

*Egzona Isufi Neziri,[a] Céline Hensky,[b,c] Hien Quy Le,[b,c] Diego Radillo Ochoa,[b,c] Aleksandra Cebrat,[a] Manfred Parschau,[a] Karl-Heinz Ernst*[\*,a,d,e] *Christian Wäckerlin*[\*,b,c]

[a] Molecular Surface Science Group, Empa, Swiss Federal Laboratories for Materials Science and Technology 8600 Dübendorf (Switzerland)

[b] Institute of Physics, Swiss Federal Institute of Technology Lausanne (EPFL) Station 3 CH-1015 Lausanne (Switzerland)

[c] Laboratory for X-ray Nanoscience and Technologies, Paul-Scherrer-Institut (PSI) CH-5232 Villigen PSI (Switzerland)

[d] Nanosurf Laboratory, Institute of Physics, The Czech Academy of Sciences 16200 Prague (Czech Republic)

[e] Department of Chemistry, University of Zurich, CH-8057 Zürich (Switzerland)

**Corresponding Authors**
\* karl-heinz.ernst@empa.ch; christian.waeckerlin@epfl.ch

## SPM imaging parameters

**Table S1.** SPM imaging parameters.

| Figure | Temperature | Imaging mode and parameters |
|---|---|---|
| **1e** | 4.5 K | Constant current<br>• Setpoint: 50 pA<br>• Bias voltage : 1000 mV |
| **1f,g** | 300 K | Constant current<br>• Setpoint: 150 pA<br>• Bias voltage : 856 mV |
| **1h,i,j** | 4.5 K | Constant height with concurrently recorded current and frequency shift<br>• Bias voltage: 1.5 mV<br>• AFM amplitude: 50 pm<br>• Current image: 4 pA (black) to 203 pA (white)<br>• Frequency shift image: −20 Hz (black) to 0 Hz (white) |

## 2D crystallinity of NiTCNE

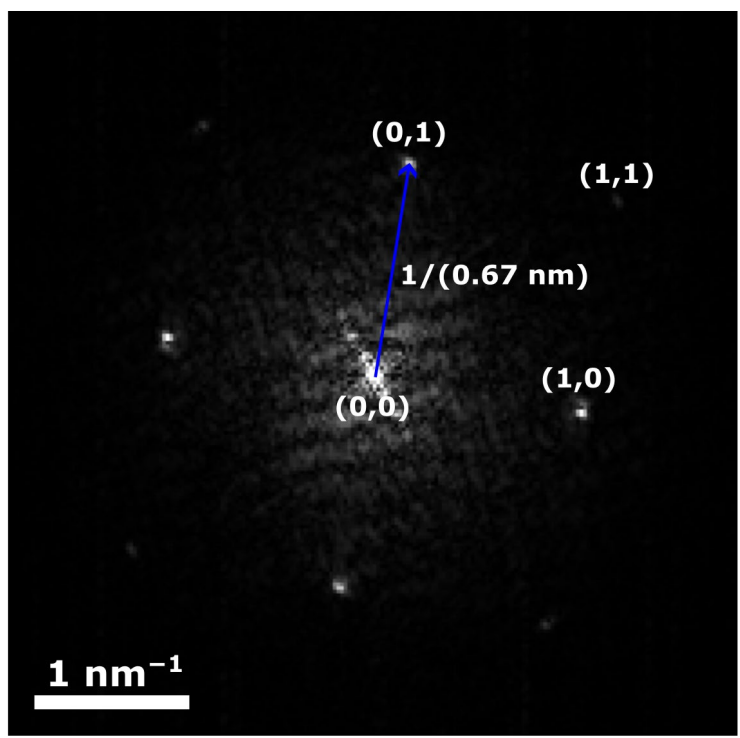

**Figure S1**. Fourier transform of the NiTCNE 2D crystal in the center of Figure 1e. The sharp peaks in reciprocal space confirm the particularly good 2D crystallinity of the lattice.

## Multiplet calculations and sum-rule analysis

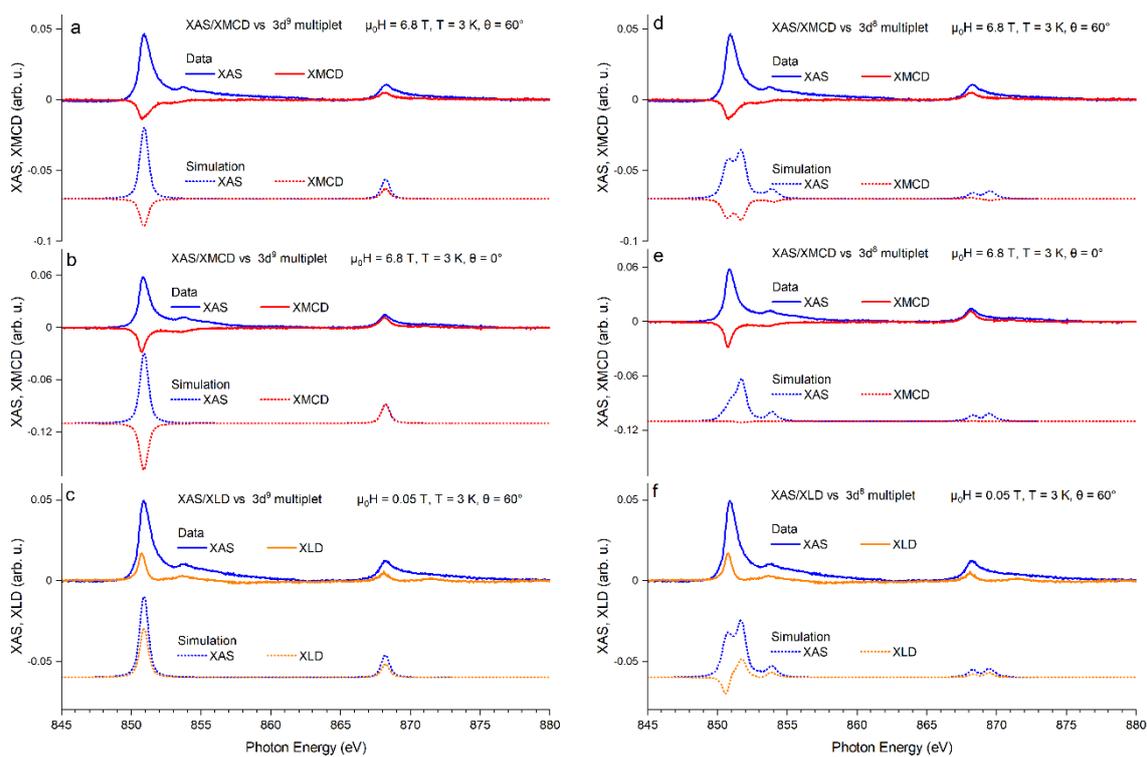

**Figure S2:** Multiplet calculations. Grazing incidence XAS/XMCD (a, d) normal incidence XAS/XMCD (b, e) and grazing incidence XAS/XLD data (c, f) of NiTCNE/Au(111) compared with simulated spectra for $3d^9$ (a-c) and $3d^8$ (d-f) electronic configurations. XAS corresponds to the sums $(\sigma_+ + \sigma_-)$ and $(\sigma_v + \sigma_h)$, respectively. XMCD and XLD correspond to the differences $(\sigma_+ - \sigma_-)$ and $(\sigma_v - \sigma_h)$, respectively.

**Table S2.** Real space crystal field coordinates used for the multiplet simulations. The calculated spectra were corrected by a constant energy offset, the value of the coulomb interaction was scaled to 80% of its computed value. The spin-orbit coupling was scaled to 95% (3d$^9$) and 91% (3d$^8$) of its computed value to reproduce the energy splitting between the L$_3$ and L$_2$ edges. The atomic coordinates were taken from DFT calculations.[1]

| X (Å) | Y (Å) | Z (Å) | Charge (e) | Atom type |
|---|---|---|---|---|
| 1.2717 | 1.5016 | 0 | -2.0 | NC |
| 2.0257 | 2.3921 | 0 | 1.0 | CN |
| 2.78 | 3.5715 | 0 | 0.5 | CC |
| 4.227 | 3.5715 | 0 | 0.5 | CC |
| 4.9813 | 2.3921 | 0 | 1.0 | CN |
| 5.7353 | 1.5016 | 0 | -2.0 | NC |
| 7.007 | 0 | 0 | 0.0 | Ni |
| 2.0257 | 4.7509 | 0 | 1.0 | CN |
| 1.2717 | 5.6414 | 0 | -2.0 | NC |
| 0 | 7.143 | 0 | 0.0 | Ni |
| 4.9813 | 4.7509 | 0 | 1.0 | CN |
| 5.7353 | 5.6414 | 0 | -2.0 | NC |
| 7.007 | 7.143 | 0 | 0.0 | Ni |
| -5.7353 | -5.6414 | 0 | -2.0 | NC |
| -4.9813 | -4.7509 | 0 | 1.0 | CN |
| -4.227 | -3.5715 | 0 | 0.5 | CC |
| -2.78 | -3.5715 | 0 | 0.5 | CC |
| -2.0257 | -4.7509 | 0 | 1.0 | CN |
| -1.2717 | -5.6414 | 0 | -2.0 | NC |
| 0 | -7.143 | 0 | 0.0 | Ni |
| -4.9813 | -2.3921 | 0 | 1.0 | CN |
| -5.7353 | -1.5016 | 0 | -2.0 | NC |
| -7.007 | 0 | 0 | 0.0 | Ni |
| -2.0257 | -2.3921 | 0 | 1.0 | CN |
| -1.2717 | -1.5016 | 0 | -2.0 | NC |
| -7.007 | -7.143 | 0 | 0.0 | Ni |
| -5.7353 | 1.5016 | 0 | -2.0 | NC |
| -4.9813 | 2.3921 | 0 | 1.0 | CN |
| -4.227 | 3.5715 | 0 | 0.5 | CC |
| -2.78 | 3.5715 | 0 | 0.5 | CC |
| -2.0257 | 2.3921 | 0 | 1.0 | CN |
| -1.2717 | 1.5016 | 0 | -2.0 | NC |
| -4.9813 | 4.7509 | 0 | 1.0 | CN |
| -5.7353 | 5.6414 | 0 | -2.0 | NC |
| -7.007 | 7.143 | 0 | 0.0 | Ni |
| -2.0257 | 4.7509 | 0 | 1.0 | CN |
| -1.2717 | 5.6414 | 0 | -2.0 | NC |
| 1.2717 | -5.6414 | 0 | -2.0 | NC |
| 2.0257 | -4.7509 | 0 | 1.0 | CN |
| 2.78 | -3.5715 | 0 | 0.5 | CC |
| 4.227 | -3.5715 | 0 | 0.5 | CC |
| 4.9813 | -4.7509 | 0 | 1.0 | CN |
| 5.7353 | -5.6414 | 0 | -2.0 | NC |
| 7.007 | -7.143 | 0 | 0.0 | Ni |
| 2.0257 | -2.3921 | 0 | 1.0 | CN |
| 1.2717 | -1.5016 | 0 | -2.0 | NC |
| 4.9813 | -2.3921 | 0 | 1.0 | CN |
| 5.7353 | -1.5016 | 0 | -2.0 | NC |

**Table S3.** Expectation values of the ground state spin and orbital moments $<S_z>$ and $<L_z>$ and of the effective spin and orbital moments $<S_{z,eff}>$ and $<L_{z,eff}>$ obtained by application of the sum-rules to the simulated spectra. The correction factor $c_S$ relates the experimentally accessible effective spin moment $<S_{eff,z}>$ obtained via sum-rule analysis to the true spin moment $<S_z>$.

| X-ray incidence angle | Ground state of multiplet | | Sum-rules applied to the calculated spectra | | Correction factor |
|---|---|---|---|---|---|
| | $<S_z>$ | $<L_z>$ | $2<S_{z,eff}>$ | $<L_{z,eff}>$ | $c_S = <S_{eff,z}> / <S_z>$ |
| 0° | 0.499 | 0.626 | 0.930 | 0.205 | 0.932 |
| 60° | 0.470 | 0.250 | 0.516 | 0.125 | 0.549 |

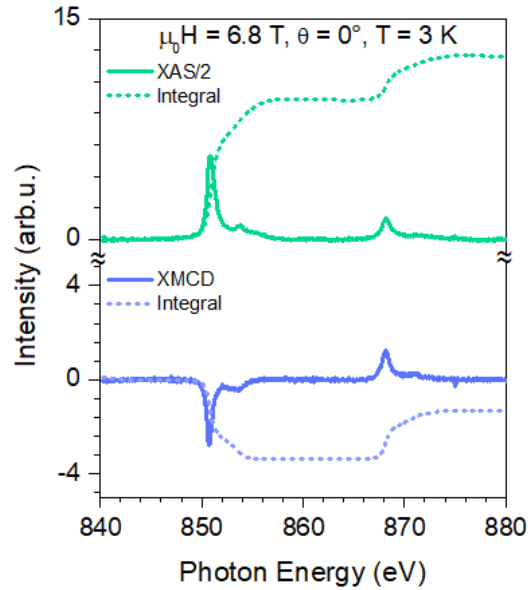

**Figure S3.** Example background subtracted x-ray spectra recorded at Ni $L_{2,3}$ edge along with the corresponding integrals relevant for the sum-rule analysis. For the sum rule analysis the isotropic absorption $XAS/2 = (\sigma_+ + \sigma_-)/2$ is used.

**Details on the sum-rule analysis and on the strongly anisotropic spin dipole moment**

Experimentally, the spin and orbital magnetic moments can be determined by application of the sum-rules[2,3] on the spectra. However, the such obtained effective spin moment $<S_{z,eff}>$ can be affected by i) spectral overlap as well as by the presence of a ii) strongly anisotropic spin dipole moment $<T_z>$.[3] Because the used multiplet code multiX considers both effects, the most straightforward approach is to apply the sum-rules to the simulated spectra, yielding $<S_{z,eff}>$ and correction factor $c_S = <S_{eff,z}> / <S_z>$ that accounts for both effects.[4] The angle dependent corrections factors are then used to correct the experimentally determined effective spin moments.

Indeed, inspection of the sum rule-results of the simulated spectra (Table S3) yields $c_S(60°) = \langle S_{eff,z}\rangle / \langle S_z\rangle = 0.549$, i.e. $\langle S_{eff,z}\rangle$ underestimates the true spin by nearly a factor of 2. In contrast, in normal incidence this factor is close to unity. Since in case of the late transition metals (Ni, Cu) the spectra overlap is insignificant,[2] the correction factor is essentially attributed to the intra-atomic spin dipole operator $\langle T_z\rangle$. Note that these findings are in very good agreement with the previously reported multiplet analysis of CuPc, a $3d^9$ case with a rather similar, square-planar crystal field[5] as well as with previously report XAS/XMCD data on NiTCNQ/Ag(100).[6]

Note that the fact that NiTCNE is fully magnetically saturated at 6.8 T in both grazing and normal incidence (Figure S7), presents the opportunity to test the validity of the above-described correction. Indeed, such obtained the experimental spin moments in normal and grazing incidence are quite similar: considering the uncertainty, the ratio $\langle S_z\rangle(0°)/\langle S_z\rangle(60°) = 0.88 \pm 0.1$ is close to unity, as expected in case of magnetic saturation.

**Supplementary XAS/XMCD data**

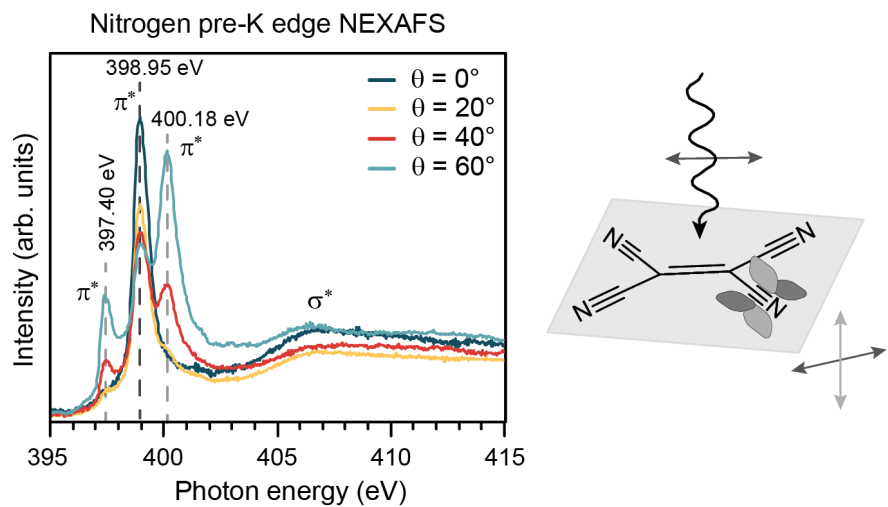

**Figure S4.** N K edge XAS of NiTCNE on Au(111) taken at incidence angles θ ranging from 0° to 60° ($\mu_0H$ = 50 mT; T = 300 K, polarization $\sigma_h$, *i.e.* nearly out-of-plane at θ = 60°). Very similar N K edge spectra were reported for NiTCNQ.[6] In principle, the split of the π* resonance can be understood by hybridization the of one of the C≡N π-bonds with the central C=C π-bond.[7,8]

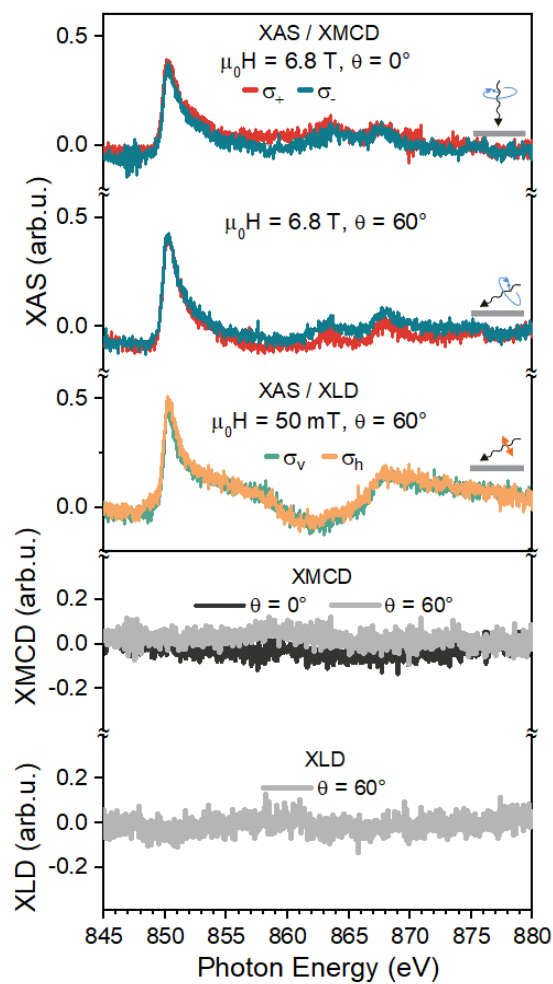

**Figure S5.** XAS/XMCD/XLD characterization of Ni clusters on Au(111). Ni clusters on Au(111) are non-magnetic, i.e. they do not exhibit any XMCD signal even at high applied magnetic field (6.8 T) and low temperature (3 K). Moreover, the XAS line-shapes are very different than the ones of NiTCNE and there is no XLD signal.

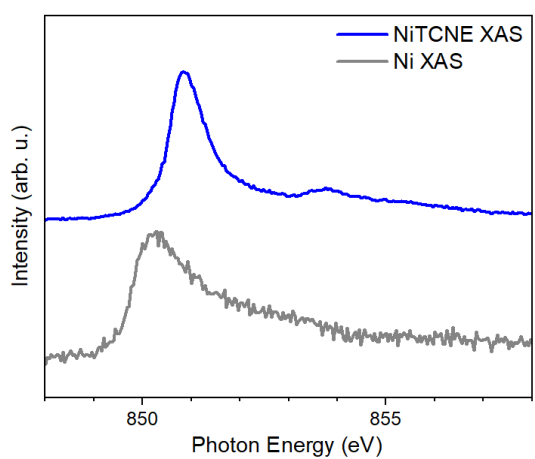

**Figure S6.** Ni L$_3$ edge XAS of NiTCNE compared with deliberately made Ni clusters on Au(111). The nickel clusters present a metallic edge-like line shape like metallic nickel, while Ni in NiTCNE exhibits a much narrower XA peak at an increased photon energy, reminiscent transition-metal elements in non-metallic form.

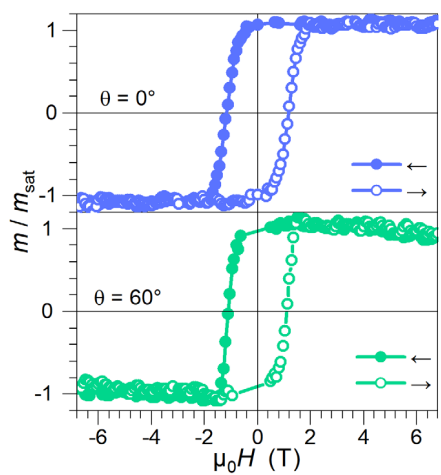

**Figure S7:** Magnetization curves of NiTCNE/Au(111) recorded at 3 K over the complete field range (±6.8 T). The field sweep rate is 2.0 T/min. The data confirms that NiTCNE is completely magnetically saturated once the hysteresis opening closes at ~ ±1 T.